\documentclass[11pt,cits]{JHEP3}

\usepackage{epsfig}
\usepackage{latexsym}
\usepackage{graphicx}
\usepackage{amsmath,amssymb}
\usepackage{epstopdf}  
\def\unit{\relax{\rm 1\kern-.26em I}}

\def\bea{\begin{eqnarray}}
\def\eea{\end{eqnarray}}
\def\be{\begin{equation}}
\def\ee{\end{equation}}
\def\nn{\nonumber}

\newcommand{\gsim}{\lower.7ex\hbox{$\;\stackrel{\textstyle>}{\sim}\;$}}
\newcommand{\lsim}{\lower.7ex\hbox{$\;\stackrel{\textstyle<}{\sim}\;$}}

\title{An Effective Description of the Landscape -- I}
\author{ \\
\\
\email{}}
\author{Diego Gallego and Marco Serone \\
International School for Advanced Studies (SISSA/ISAS)
and INFN, Trieste, Italy\\
E-Mail: \email{gallego@sissa.it,serone@sissa.it}}

\abstract{We study under what conditions massive fields can be  ``frozen''  rather than integrated out 
in  certain four dimensional theories
with global or local  ${\cal N}=1$ supersymmetry.
We focus on models without gauge fields, admitting a superpotential of the form 
$W = W_0(H) + \epsilon\, W_1(H,L)$, with $\epsilon \ll 1$, where $H$ and $L$ schematically
denote the heavy and light  chiral superfields. 
We find that the fields $H$ can always be frozen to constant values $H_0$,  if they approximately correspond to  
supersymmetric solutions along the $H$ directions, 
independently of the form of the K\"ahler potential $K$ for $H$ and $L$,
provided $K$ is sufficiently regular. 
In supergravity $W_0$ is required to be of order $\epsilon$ at the vacuum
to ensure a mass hierarchy between $H$ and $L$.
The backreaction induced by the breaking of supersymmetry
on the heavy fields is always negligible, leading to suppressed $F^H$--terms.
For factorizable K\"ahler potentials $W_0$ can instead be generic.
Our results imply that the common way complex structure and dilaton moduli are stabilized,
as in  Phys.\ Rev.\  D {\bf 68} (2003) 046005 by Kachru et al.,  for instance, is reliable to a very good accuracy, provided
$W_0$ is small enough.}

\preprint{SISSA-72/2008/EP}


\begin{document}

\section{Introduction}

Effective field theories are probably the main tool of investigation in particle physics.
Thanks to the decoupling of massive particles in low energy processes at the quantum level \cite{Appelquist:1974tg}, 
effective theories provide a significant and reliable simplification in the description of most physical systems.
In a standard quantum field theory perturbative context, where the location of the vacuum and 
the structure of the tree level Lagrangian is typically simple to determine, one commonly integrates out heavy fields  
to get an effective field theory where processes at the classical and quantum level are efficiently computed.

The situation is a bit different in string theory, where the number of degrees of freedom is so large (actually infinite), that
massive particles are typically neglected, rather than integrated out. At a first approximation this is fine, provided such states
are easily detected and sufficiently heavy, which is the case for string excitations or Kaluza--Klein states in a
given compactification.   Below the string or compactification scale, the remaining ``light'' states are typically described by a given supergravity (SUGRA) field theory, assumed from now on to be four-dimensional with ${\cal N}=1$ supersymmetry (SUSY). Despite this huge simplification, the resulting SUGRA field theory  typically contains hundreds of fields, making an explicit  full study of these theories a formidable task.
Thanks to the  recent progress in string compactifications with fluxes,
many of these fields,  which are moduli fields in an adiabatic approximation where fluxes are turned off, can get
large masses and one can hope to get rid of them, in the spirit of effective theories.
The scalar potential associated to these theories is however very complicated, admitting in general a ``landscape'' of possible vacua \cite{Susskind:2003kw}, and even just determining the location of the vacua is often very hard.  
As a matter of fact, in most cases a large subset of the moduli are neglected  without a proper justification.
They are assumed to get a sufficiently large mass by a tree-level flux-induced superpotential or somehow have suppressed couplings with the remaining light fields, whereas the general correct procedure would consist in integrating them out
from the action and then check whether the generated effective couplings are negligible or not.

In simple global SUSY theories a heavy chiral superfield $H$ is easily integrated out at tree-level and at low energies (when derivatives interactions are negligible) in unbroken SUSY vacua, by just taking $\partial_H W =0$
and plugging back the solution in the superpotential $W$  and the K\"ahler potential $K$. In this way one gets an effective holomorphic superpotential and K\"ahler potential for the remaining fields, see e.g. \cite{Affleck:1984xz}.\footnote{Ref.~\cite{deAlwis:2005tg} found that the equation $\partial_H W =0$ in general receives corrections. We believe that the corrections found in  \cite{deAlwis:2005tg} are automatically
taken into account in substituting the solution for $\partial_H W =0$ back in the K\"ahler potential as well, in which case $\partial_H W =0$ is correct and always leads to the correct effective potential.}
 This procedure is easily implemented in the SUSY theories with a simple vacuum structure, but not in the complicated SUGRA theories we are interested in, since it assumes the knowledge of the vacuum and of the mass spectrum, which are
actually the first quantities to be determined. Even if we have some guess about which are the heavy fields $H$ in a given region in moduli space of the theory, their Vacuum Expectation Values (VEV's) will in general depend on the remaining light fields $L$, so that already at the vacuum level we encounter problems to decouple heavy and light degrees of freedom.
In this situation, it is not well defined the notion of  ``freezing''  the heavy moduli $H$, namely fixing them at their VEV's and
neglecting their quantum fluctuations, and practically impossible to properly integrate them out. 
 As a further complication, integrating out $H$ in SUGRA generally would give rise to a non-holomorphic and quite complicated effective superpotential for the light fields $L$, as emphasized in \cite{deAlwis:2005tf}. The non-holomorphicity of the superpotential is not in contradiction with SUSY, since the splitting between $K$ and $W$ in SUGRA is arbitrary, being $G=K+\log |W|^2$, the only K\"ahler invariant quantity,  but certainly the resulting effective theory will generally be quite awkward. 

Aim of the present paper is to study under what conditions massive fields can be frozen rather than properly integrated out, and yet get a reliable effective theory, in 4D theories with ${\cal N}=1$  global or local supersymmetries. We focus on this paper
on theories with only chiral multiplets and no gauge fields. In a companion paper, we will extend the present results 
to the more interesting, but more involved, case where gauge fields are added \cite{EffD}.
As mentioned before, a meaningful freezing of the heavy fields requires that in some approximation
their VEV's  do not depend on the light fields and can be fixed to constant values $H_0$, independently of
the light field dynamics. We are then led to study theories with a superpotential of the form 
\be
W(H,L) = W_0(H) + \epsilon \, W_1(H,L)\,, 
\label{Wint}
\ee
with $\epsilon \ll 1$, where $H$ and $L$ schematically
denote heavy and light  chiral superfields.\footnote{This distinction is independent of the K\"ahler potential. 
As we will see, it is always possible to define a canonically normalized field basis where the fields $H$ give rise to the canonical
heavy field fluctuations.}
The K\"ahler potential is arbitrary, with the only assumption that all the eigenvalues
of the associated K\"ahler metric are parametrically larger than $\epsilon$. By expanding the theory in a series in $\epsilon$, 
we find that for theories with global SUSY the fields $H$ can always be reliably frozen to constant values $H_0$ (defined as the approximate SUSY solutions $F_{H,0} = \partial_H W_0 = 0$) 
independently of the form of the K\"ahler potential $K$ for $H$ and $L$.
By reliably, we mean that the {\it complete}  bosonic effective low-energy theory obtained by freezing the fields $H$ 
is identical, at leading order in $\epsilon$, to the full bosonic effective theory obtained by classically integrating $H$ out at the gaussian level at low energies, i.e. neglecting their kinetic terms.
In the following, for simplicity of language, we will denote the former and latter ``simple" and ``full" effective theories, respectively.   This result applies on any vacuum, supersymmetric or not.  In SUGRA theories the covariant derivative
entering in the definition of the $F$ term would lead to the condition
$F_{H,0} = \partial_H W_0+(\partial_H K) W_0 = 0$, rather than $\partial_H W_0 = 0$.
The former equation depends in general on the light fields through $K$ and does
not  allow for a well--defined ``freezing'' for the fields $H$. An extra requirement is hence needed.
Either i)  the K\"ahler potential approximately factorizes, $K=K_H(H)+K_L(L)+O(\epsilon)$, so that
$F_{H,0}=0$ does not depend on $L$ at leading order in $\epsilon$ or ii)
$W_0(H_0)\lesssim O(\epsilon)$, so that effectively $F_{H,0} = \partial_H W_0 =0$.
A factorisable K\"ahler potential, together with a superpotential of the form (\ref{Wint}), fulfills actually 
a  sufficient condition for decoupling found in \cite{Binetruy:2004hh} (see also \cite{Achucarro:2007qa}).  
For this reason, we mainly focus on the second possibility, where the K\"ahler potential is generic. The smallness of $W_0(H_0)$ is well justified from the point of view of an effective theory of light fields, given the rather universal SUGRA contributions to the masses, proportional to $\langle W\rangle$. From a phenomenological point of view, this is related to the requirement of low energy spontaneous supersymmetry breaking and a not too large gravitino mass, given by 
\be
m_{3/2}=e^{K/2} |W|\,.
\ee
We are eventually interested in vacua where SUSY is spontaneously broken, so another important effect that has to be considered is the backreaction of the SUSY breaking sector on the heavy fields.  Eq.(\ref{Wint}) and $F_{H,0}=0$ generally predicts 
\be
F_L \sim F_H \sim O(\epsilon)\,.
\ee
Interestingly enough,  independently of the form of $K$, the upper $F$ term component 
are
\be
F^L \sim O(\epsilon)\,, \ \ \ \ F^H \sim O(\epsilon^2)\,,
\ee
so that at leading order the backreaction of SUSY breaking on the heavy sector is negligible, in agreement
with the result that the fields $H$ can reliably be frozen at an approximately SUSY solution along the $H$ directions.\footnote{This can also be seen by noting that the SUSY transformations of the heavy fermions are proportional to $F^H$.}
These results are valid for any value of the cosmological constant.

A natural class of string models where our general results can be applied are IIB/F-theory Calabi-Yau compactifications with fluxes, where $W_0$ is identified with the Gukov-Vafa-Witten flux superpotential \cite{Gukov:1999ya}, generally able to fix all complex structure moduli, including the dilaton \cite{Giddings:2001yu}.  Fixing complex structure moduli is actually the first step of the Kachru-Kallosh-Linde-Trivedi (KKLT) procedure \cite{Kachru:2003aw} in finding de Sitter (dS) SUGRA vacua with stabilized moduli in string theory. The complex structure moduli and the dilaton can  collectively be identified with our fields $H$. The K\"ahler structure moduli, as well as other fields possibly responsible for SUSY breaking, can be identified with the fields $L$.  In this case, $W_1(L,H)$ includes non-perturbatively generated
superpotential terms for the K\"ahler moduli. Our results prove that $H$ can naively and reliably be frozen by just using $W_0$, as  \cite{Kachru:2003aw} does.  One can then consistently forget about the complex structure and dilaton dependence appearing in $W_1$, possible K\"ahler mixing between $H$ and $L$, which do typically appear 
at some order in $\alpha^\prime$ or $g_s$, and about how the fields $H$ contribute to the SUSY breaking mechanism induced by the fields $L$, provided $W_0\sim \epsilon$ at the vacuum.\footnote{If the K\"ahler potential happens to be almost factorizable, the condition $W_0\sim \epsilon$ might be partially relaxed, since in this case a sort of mixing between the decoupling effects i) and ii) discussed before apply.}
In \cite{Kachru:2003aw} the tuning $W_0\ll 1$ was actually required for another reason, namely to supersymmetrically stabilize the universal K\"ahler modulus to large values. We find here that the smallness of $W_0$ is more important than previously
thought, being the basis for a consistent decoupling of the two moduli sectors. 
One has also to check  that the mass matrix of the $H$ fields, governed by $W_0$, is positive definite with all eigenvalues parametrically larger than the effective $\epsilon$ defined by $W_1$.\footnote{See also \cite{Abe:2006xi} where the condition $W_0\ll 1$ has been shown to be necessary to have a reliable description of the vacuum in the effective theory.} 
As a matter of fact, previous works pointing out problems in the way KKLT stabilize complex structure and dilaton moduli
boil down to not satisfy either the requirement $W_0\ll 1$ \cite{deAlwis:2005tf}, or the requirement that all the heavy moduli
get large positive SUSY mass terms \cite{Choi:2004sx}. Interestingly enough, one typically requires many complex structure
moduli to be able to tune $W_0$ and the cosmological constant \cite{Bousso:2000xa} to sufficiently small values, in which case there is no reason to expect small or negative
SUSY mass terms \cite{Hebecker:2006bn,Denef:2004ze}.\footnote{We thank S. Kachru for bringing \cite{Hebecker:2006bn} to our attention.}  
It is important to remark that our results are valid in the context of a spontaneous SUSY breaking mechanism. In this sense, they cannot automatically provide a solid framework for the whole KKLT procedure, since it is not yet clear whether the SUSY breaking mechanism advocated in \cite{Kachru:2003aw} (the addition of $\bar D_3$-branes) admits an interpretation in terms of a spontaneously
broken 4D ${\cal N}=1$ SUGRA theory.

Although we mostly focus on SUGRA models with $W_0\sim  \epsilon$, we also briefly study how eq.(\ref{Wint})  leads to decoupling,  when $W_0$ is generic, but  $K$ is almost factorizable. We then consider in some detail a relevant model of this sort,  namely a so called large volume model \cite{Balasubramanian:2005zx,Conlon:2005ki} arising from flux compactifications of Type IIB string theory on Calabi-Yau (CY) 3-folds with an exponentially large volume.  The superpotential associated to this model is effectively of the form (\ref{Wint}), with $\epsilon$ related to the inverse volume of the CY. There is not a mass hierarchy between the heavy and light fields, but nevertheless the couplings between the two sectors are suppressed, so that an effective decoupling occurs, namely the full and simple effective potentials agree at leading order in $\epsilon$. 

The structure of the paper is as follows. In section 2 we introduce the basic elementary tools used throughout the paper 
in a simple non-supersymmetric bosonic $\sigma$-model. In section 3 we derive 
the equivalence between the full and the simple effective actions for a generic SUSY model in flat space and with a superpotential of the form (\ref{Wint}). In section 4 we show how the analysis straightforwardly applies
to SUGRA, provided $W_0$ is small at the minimum for a generic K\"ahler potential or the latter is almost factorizable 
and $W_0$ generic. In section 5 we conclude.  We explicitly work out in three Appendices some concrete models for illustrative purposes. In Appendix A we study a very simple model with global SUSY, in Appendix B a more complicated KKLT--like model and finally, in Appendix C, we report the study of a large volume model as a specific example of a SUGRA model with an almost factorizable K\"ahler potential.

\section{Non-Supersymmetric $\sigma$-Model}

Before considering the more complicated SUSY case, we establish a very simple, yet useful, result valid for 
an arbitrary non-supersymmetric bosonic $\sigma$-model. 
Let us consider a system of $n_H+n_L$ interacting real scalar fields $H^i$ and $L^\alpha$, $i=1,\ldots,n_H$, $\alpha=1,\ldots,n_L,$ with Lagrangian density
\be
{\cal L} = \frac 12 g_{MN}(\phi^M) \partial \phi^M \partial \phi^N - V(\phi^M)\,,
\label{LnoSUSY}
\ee
and potential
\be
V(\phi^M) = V_0(H^i) + \epsilon \, V_1 (H^i,L^\alpha)\,,
\ee
where $M=1,\ldots,n_H,1+n_H,\ldots,n_H+n_L$, $\phi^M=(H^i,L^\alpha)$ and $\epsilon\ll 1$. 
Notice that the splitting between the fields $H^i$ and $L^\alpha$ is dictated by $V_0$,
namely we call $H^i$ (the ``heavy fields") the ones appearing in $V_0$. We assume that at a given vacuum $\langle \phi^M\rangle $, the metric $g_{MN}$ is non-singular and $\partial_i \partial_j V_0$ is positive definite with all eigenvalues
parametrically larger than $\epsilon$.  Under these assumptions, at leading order in $\epsilon$, the simple effective low-energy Lagrangian
associated to ${\cal L}$ is 
\be 
{\cal L}_{sim} =  \frac 12 g_{\alpha \beta}(L^\alpha,H_0^i) \partial L^\alpha \partial L^\beta - \Big[ V_0(H_0^i)+\epsilon \, V_1(H_0^i,L^\alpha)\Big]\,,
\label{LeffnoSUSY}
\ee
where $H_0^i$ are the leading order VEV's for $H^i$, independent of $L^\alpha$, satisfying $\partial_i V_0(H_0^j)=0$.
We want to show that ${\cal L}_{sim}$ provides the correct effective Lagrangian for arbitrary kinetic mixing terms at leading order in an expansion in $\epsilon$.
First of all, we look for (space-time independent) vacua of ${\cal L}$ by studying the extrema of the potential $V$ in the original, non-canonically normalized, field basis given by $H^i$ and $L^\alpha$. Let $H_0^i$ be the solutions to $\partial_i V_0 = 0$  at $O(\epsilon^0)$. At this order, $L^\alpha$ are  undetermined, since $\partial_\alpha V_0$ trivially vanishes. At $O(\epsilon)$, we get the leading VEV's $L_0^\alpha$ for the light fields from $\partial_\alpha V_1(H_0^i) = 0$ and the first corrections $H_1^i$ to the VEV's of the heavy fields from the linear equations $\partial_i \partial_j V_0(H_0^k) H_1^j +\partial_i V_1(H_0^k,L_0^\alpha) = 0$. Once the approximate vacuum $(H_0^i,L_0^\alpha)$ has been determined,  we can diagonalize the metric in order to identify the canonically normalized field fluctuations $\hat \phi_c$ starting from the field fluctuations $\hat \phi=\phi-\langle\phi\rangle$.\footnote{We thank A. Romanino for essentially providing us the argument that follows.}
 At leading order the matrix to diagonalize is $g_{MN,0}=g_{MN}(H_0^i,L_0^\alpha)$ and, for simplicity, we assume all its eigenvalues to be parametrically larger than $\epsilon$. A positive definite real symmetric matrix can always 
be written as the product of a lower triangular matrix times its transpose.\footnote{This procedure, called Cholesky  decomposition, is unique if we require the diagonal entries of $T$ to be strictly positive. See e.g. \cite{golub}.}
 We then write
$g_0 =  (T^{-1})^t T^{-1}$, so that $\hat\phi = T \hat \phi_c$, where 
\be
T = \left(
\begin{matrix}
(T_H)^i_j &  0  \cr
 (T_{HL})^\alpha_i & (T_L)^\alpha_\beta  \cr
\end{matrix}\;
\right)\label{Ttr}
\ee
and $\hat \phi=(\hat H^i,\hat L^\alpha)^t$.
In this new basis, the Lagrangian (\ref{LnoSUSY}) reads
\bea
\!\!\!\!\!{\cal L} = &&\Big[\frac12+O(\epsilon) \Big] \Big[(\partial \hat H_c^i)^2 + (\partial \hat L_c^\alpha)^2 \Big] +\ldots +\nn \\
\!\!\!\!\! &&\!\! \Big[ V_0(H_0^i+\epsilon H_1^i+(T_H \hat H_c)^i)
+\epsilon V_1(H_0^i+\epsilon H_1^i+(T_H \hat H_c)^i, L_0^\alpha+ (T \hat \phi_c)^\alpha)+O(\epsilon^2)\Big],
\label{VnoSUSY2}
\eea
where the ellipsis in eq.(\ref{VnoSUSY2}) stands for all the higher order terms arising from the expansion of the metric
in quantum fluctuations, the specific form of which are not needed. 
Thanks to the triangular form of $T$,  $\hat H_c^i$ are linear combinations of the $\hat H^i$ only. 
The next step would be to diagonalize the mass matrix of the heavy fields, but it will not be explicitly needed.
Indeed, we see from the term $V_0$ in eq.(\ref{VnoSUSY2}) that the fields $\hat H_c^i$  have all a leading mass term of $O(\epsilon^0)$,
the $n_H\times n_H$ mass matrix being of the form  $M^2=T_H^t M_0^2 T_H$, where $(M_0^2)_{ij} = \partial_i\partial_j V_0(H_0^i)$. Since by assumption $M_0^2$
is positive definite, so it is $M^2$.
Due to the form of the potential in eq.(\ref{VnoSUSY2}), integrating out the fluctuations $\hat H_c^i$ at quadratic level will only affect the effective theory at $O(\epsilon^2)$, so we can simply set $\hat H_c^i = 0$ if we want a reliable Lagrangian up to $O(\epsilon)$. We can go back to non-canonically normalized fields $\hat  L^\alpha= (T_L^{-1} \hat L_c)^\alpha$. 
Since $T_L^{-1} = (T^{-1})^\alpha_\beta$, the kinetic mixing matrix reads now $(T_L^{-1})^t (T_L)^{-1} =  g_{\alpha\beta,0}$
and all the terms in the ellipsis in eq.(\ref{VnoSUSY2}), when $\hat H_c^i = \hat H^i=0$, simply reproduce the full field-dependent metric $g_{\alpha\beta}(L^\alpha,H_0^i)$ appearing in eq.(\ref{LeffnoSUSY}).
The resulting full effective Lagrangian becomes 
\be
{\cal L}_{full} = \Big[\frac12 g_{\alpha\beta}(L^\alpha,H_0^i) +O(\epsilon) \Big] \ \partial  L^\alpha \partial L^\beta  - \Big[ V_0(H_0^i+\epsilon H_1^i)
+\epsilon V_1(H_0^i, L^\alpha)+O(\epsilon^2)\Big] \nn
\label{LeffnoSUSY2}
\ee
and hence, modulo irrelevant constant terms, at leading order in $\epsilon$  we get the desired result
\be
{\cal L}_{sim} = {\cal L}_{full}\,.
\ee

\section{Supersymmetric $\sigma$-Model}

The results found in the previous section, being of general validity, apply to SUSY theories as well.
But the structure of the scalar potential fixed by SUSY allows for a more powerful result, valid up to $O(\epsilon^2)$.
Neglecting gauge fields, a SUSY theory is specified by a K\"ahler potential $K(\phi, \bar \phi)$, taken to be generic,
and a superpotential, taken as in eq.(\ref{Wint}):
\be
W(H^i,L^\alpha) = W_0(H^i) +\epsilon W_1(H^i,L^\alpha)\,,
\label{WSUSY}
\ee 
using the same conventions as before, but considering that now $H^i$ and $L^\alpha$\ are complex (super)fields.\footnote{For simplicity of notation, below and throughout the paper, we use the same
notation to denote a chiral superfield and its lowest scalar
component, since it should be clear from the context to what we
are referring to. Similarly, we will mostly not report the complex conjugate VEV's of the fields, their corresponding equations of motion, and so on.}  The scalar potential of the theory is
\be
V = g^{\bar M N} \overline F_{\bar M} F_N\,,
\label{VSUSY}
\ee
with $F_M = \partial_M W$ and $g^{\bar M N}$ the inverse matrix of $g_{M\bar N} = \partial_M\partial_{\bar N}K$. 
The potential $V$ is a sum of three terms when expanded in $\epsilon$: 
$V = V_0 + \epsilon V_1 + \epsilon^2 V_2$, with
\bea
V_0 & = &  g^{\bar j i }\overline F_{\bar j,0} F_{i,0} \,, \nn \\
V_1 & = & g^{\bar j i}   \overline F_{\bar j,1} F_{i,0} + g^{\bar \alpha i}\overline F_{\bar \alpha,1}  F_{i,0} + c.c. \,,
\label{V012} \\
V_2 & = & g^{\bar j i}  \overline F_{\bar j,1} F_{i,1}   + g^{\bar \beta \alpha}   \overline F_{\bar \beta,1} F_{\alpha,1} + 
(g^{\bar \alpha i}   \overline F_{\bar \alpha,1}  F_{i,1}  + c.c. ) \,, \nn
\eea
where $F_M = F_{M,0}+\epsilon F_{M,1}$ and
\bea
F_{i,0} & = &  \partial_i W_0 \,, \hspace{1cm}  F_{\alpha,0}  =  0\,, \\
F_{i,1} &  = &   \partial_i W_1\,, \hspace{1cm} F_{\alpha,1} =  \partial_\alpha W_1\,. \nn
\eea
We assume that at the vacuum the metric $g_{M\bar N}$ is positive definite and that $\partial_i \partial_j W_0$  is non-degenerate with eigenvalues parametrically larger than 
$\epsilon$.
We now establish that if the heavy fields sit at a SUSY vacuum at leading order in $\epsilon$, the bosonic low energy effective theory of the light fields $L^\alpha$ is described by the simple SUSY effective theory with 
$K_{sim}(L^\alpha,\bar L^\alpha) = K(H_0^i,L^\alpha,\bar H_0^i, \bar L^\alpha)$, $W_{sim} = W_0(H_0^i)+ \epsilon W_1(H_0^i,L^\alpha)$:
\be
{\cal L}_{sim} = \tilde g_{ \alpha \bar \alpha} \partial L^\alpha \partial \bar L^{\bar \alpha} - V_{sim}(L^\alpha,\bar L^{\bar \alpha})\,,
\label{Lsim}
\ee
with $V_{sim} = \tilde g^{\bar \alpha \alpha} \tilde F_{\alpha} \overline{\tilde F}_{\bar \alpha}$,
$\tilde g_{\alpha \bar \alpha} = \partial_\alpha \partial_{\bar \alpha} K_{sim}$ and  $\tilde F_{\alpha} =\partial_\alpha W_{sim}$.

Let us start by finding the vacuum in an expansion in $\epsilon$: 
\bea
\langle \phi^M \rangle = \phi_0^M + \epsilon \phi_1^M + \epsilon^2 \phi_2^M + \ldots \,.
\eea
The equations of motion (e.o.m.) up to $O(\epsilon^2) $ read
\bea
\!\!\!\!\!\!(\partial_M V)_0 & = &  \partial_M V_0 =0\,,  \label{eqV0} \\
\!\!\!\!\!\!(\partial_M V)_1 & = &  \partial_M V_1+ (\partial_M \partial_N V_0) \phi_1^N +
(\partial_M \partial_{\bar N} V_0) \bar\phi_1^{\bar N} =0  \,,   \label{eqV1}  \\
\!\!\!\!\!\!(\partial_M V)_2 & = &  \partial_M V_2+ (\partial_M \partial_N V_1) \phi_1^N +
(\partial_M \partial_{\bar N} V_1) \bar\phi_1^{\bar N} +  (\partial_M\partial_N V_0) \phi_2^N +
  (\partial_M\partial_{\bar N} V_0) \bar\phi_2^{\bar N} \nn \\
   &+&\!\!\frac 12  (\partial_M\partial_N \partial_P V_0) \phi_1^N \phi_1^P +\frac 12 
  (\partial_M\partial_{\bar N} \partial_{\bar P} V_0) \bar\phi_1^{\bar N} \bar\phi_1^{\bar P}
 +  (\partial_M\partial_{N} \partial_{\bar P} V_0) \phi_1^N \phi_1^{\bar P}
=0 , \label{eqV2} 
\eea
where all quantities in eqs.(\ref{eqV0}), (\ref{eqV1}) and (\ref{eqV2})  are evaluated at $\phi^M=\phi_0^M$.
 At $O(\epsilon^0)$, the equations $(\partial_k V)_0 =0$ \
can generally admit both SUSY and non-SUSY solutions. Contrary to the latter ones, which can depend on $L^\alpha$ through the inverse metric components $g^{\bar j i}$,  the SUSY solutions depend on $H^i$ only. Let $H_0^i$ be the SUSY solutions:
\be
F_{i,0}(H_0^i) =\overline F_{\bar i,0 }(\bar H_0^i)  = 0\,.
\ee
The e.o.m. $(\partial_\alpha V)_0=0$ are identically satisfied when $H^i = H_0^i$, so that $L^\alpha$ are not determined at this order.
Interestingly enough,  the e.o.m. of the light fields  at $O(\epsilon)$, $(\partial_\alpha V)_1=0$, 
are automatically satisfied  as well, since
\be
\partial_\alpha \partial_M V_0 = \partial_\alpha \partial_{\bar M} V_0= \partial_\alpha V_1 = 0\,,
\label{derV0}
\ee
when evaluated at $H=H_0^i$,  as can easily be checked using the explicit expressions in eq.(\ref{V012}). The displacement of the heavy field VEV's at $O(\epsilon)$ is calculated by taking $M=j$ in eq.(\ref{eqV1}). We get
\be
H_1^i = - (K^{-1})^i_{\bar j} \overline F^{\bar j}_1\,,
\label{PhiH1}
\ee 
with $K^{\bar i}_j  = g^{\bar i k}\partial_k \partial_j W_0$ and $F_1^i = \overline F_{\bar M,1} g^{\bar M i} $, evaluated again at $H^i=H_0^i$.
Notice that $H_1^i$ are in general a function of both $L^\alpha$ {\it and} $\bar L^{\bar\alpha}$, which are yet to be determined at this order. Hence, up to  $O(\epsilon)$,  eq.(\ref{PhiH1}) is not only valid at the vacuum but is a field identity\footnote{Since we are neglecting gradient term contributions arising when $H_1$ is space-time dependent, eq.(\ref{PhiH1x}) is only valid at low energies, when $L^\alpha$ is slowly varying, which is the case of interest.} 
\be
H_1^i(x) =  - (K^{-1})^i_{\bar j} \overline F^{\bar j}_1(H_0^i,L^\alpha(x))\,.
\label{PhiH1x}
\ee
This is a crucial property to quickly establish the equivalence of the full and simple effective actions, as we will show. 
At $O(\epsilon^2)$ we finally get non-trivial e.o.m. for $L^\alpha$ as well as the $2^{nd}$ order displacement of the heavy fields $H_2^i$, whose explicit form will not be needed. 
Both $H_2^i$ and $L_0^\alpha$ arise at $O(\epsilon^2)$, but the e.o.m. of $L^\alpha$ do not depend on $H_2^i$, as can be seen from eqs.(\ref{eqV2}) and (\ref{derV0}). 
By plugging eq.(\ref{PhiH1}) in eq.(\ref{eqV2}) and after some algebra, we could establish that the e.o.m. that determine $L_0^\alpha$ in the full theory are the same 
as the one obtained in the simple theory where $H^i$ are frozen to their leading values $H_0^i$.
The best and more instructive way to proceed, however, is by finding the leading power in $\epsilon$ of the $F$ terms and their derivatives, evaluated at the shifted vacuum $H_0^i+\epsilon H_1^i$. We get
\bea
F_i & = &  O(\epsilon), \hspace{.8cm} F_\alpha = O(\epsilon), \hspace{.7cm} F^i = O(\epsilon^2), 
\hspace{.6cm}  F^\alpha = O(\epsilon)\,, \nn  \\
\partial_j F_i &  = & O(1), \ \ \ \partial_i F_\alpha = O(\epsilon), \ \ \ 
\partial_\beta F_i  =  O(\epsilon),  \ \ \ \partial_\beta F_\alpha = O(\epsilon)\,.  \label{FdF} 
\eea
Interestingly enough, although in general the backreaction of the light fields on the heavy ones induce $F_i$--terms of  $O(\epsilon)$,
the upper components $F^i$ are vanishing at this order since
\be
\overline F^{\bar i}(H_0^i+\epsilon H_1^i) = \Big[g^{\bar i M} F_{M,1}(H_0^i) + g^{\bar i j} \partial_k F_{j,0}  H_1^k \Big] \epsilon  + O (\epsilon^2)
\label{Fiup}
\ee 
and the first two terms in eq.(\ref{Fiup}) exactly cancel, due to eq.(\ref{PhiH1}). 
This implies that at linear order the $F_i$ and $F_\alpha$ terms are related as follows:
\be
F_i = - \tilde g_{i\bar j} g^{\bar j \alpha} F_{\alpha}+O(\epsilon^2)\,,
\label{FiFarel}
\ee
with $\tilde g_{i \bar j}$ the inverse metric of $g^{\bar j i}$, not to be confused with $g_{i \bar j}$.
Using eq.(\ref{Fiup}) and the relation (\ref{FiFarel}), after some straightforward algebra one finds the
desired identification 
\be
\partial_\alpha V = \partial_\alpha V_{sim}+O(\epsilon^3),
\label{VeffeqV}
\ee
which implies that the location of the vacuum is reliably computed in the simple theory. 
In order to establish eq.(\ref{VeffeqV}), it is very useful to use the matrix identity
\be
\tilde g^{\bar \alpha \alpha} = g^{\bar \alpha \alpha} - g^{\bar \alpha i} \tilde g_{i\bar j}g^{\bar j \alpha}\,,
\label{MatrixId}
\ee
where $\tilde g^{\bar \alpha \alpha}$ is the inverse of $g_{\alpha \bar \alpha}$, appearing in $V_{sim}$,
not to be confused with $g^{\bar \alpha \alpha}$. Thanks to eq.(\ref{MatrixId}), in particular, it is easy to show that $F^\alpha = \tilde F^\alpha$
at leading order in $\epsilon$.

The full leading order equivalence of the naive low-energy effective theory with the full one proceeds along the same lines
of the general non-SUSY case discussed before.  
Once found the leading order VEV's $\phi_0^M$,  we write 
$g_0 =  (T^{-1})^\dagger T^{-1}$ (which is the generalization of the Cholesky decomposition for hermitian matrices)
and get the canonically normalized field fluctuations $\hat \phi_c^M$ as $\hat\phi = T \hat \phi_c$,  $\hat{\bar\phi} = T^* \hat{\bar  \phi}_c$.  At quadratic order in the fluctuations,  the leading potential term $V_0$ in eq.(\ref{V012})  depends on $\hat H_c^i$ only. The latter are hence identified as the proper heavy fields fluctuations, with a mass matrix $T_H^\dagger M^2 T_H$ and
\be
M^2_{\bar i  j}  =  \partial_{\bar i} \partial_{\bar k} \overline W_0
g_0^{\bar k l} \partial_l \partial_j  W_0\,.
\label{Mass0}
\ee 
By assumption, the mass matrix (\ref{Mass0}) has eigenvalues parametrically larger than $\epsilon$. 
We can expand the scalar potential up to quadratic order in $\hat H_c^i$ and hence integrate the heavy
fluctuations out. Schematically the expansion is as follows
\be
V(H,L) = V(\langle H\rangle,L) + \partial_H V(\langle H\rangle,L) T_H \hat H_c +   \partial_H^2 V(\langle H\rangle,L) (T_H \hat H_c)^2 + O(\hat H_c^3)\,.
\label{SchematicExp}
\ee
Since $\partial_H V$ is at most of  $O(\epsilon)$ and $\partial_H^2 V$ of $O(1)$, $\hat H_c$ is  $O(\epsilon)$. In order to get the effective Lagrangian up to $O(\epsilon^2)$, 
it is enough to determine the $O(1)$ terms of $\partial_H^2 V$ and the $O(\epsilon)$ terms of
$\partial_H V$. Using eqs.(\ref{FdF}) and (\ref{FiFarel}), it is straightforward to show
that the linear term in $\hat H_c^i$ vanishes, being proportional to $F^i = O(\epsilon^2)$.
Hence, integrating out $\hat H_c^i$ just amounts to set them to zero!
The full effective scalar potential for the light fields $\hat L_c^\alpha$ is easily determined.  
Both $V_0(\hat L_c^\alpha)$ and $V_1(\hat L_c^\alpha)$ vanish, since $F_{i,0}(\hat H_c^i=0)=0$. 
Using eqs.(\ref{FiFarel}) and (\ref{MatrixId}) in $V_2$ in eq.(\ref{V012}), one immediately gets
\be
V  =  \epsilon^2 \tilde g^{\bar \alpha \alpha}   \overline F_{\bar \alpha,1} F_{\alpha,1}\,.
\ee
Finally, going back to non-canonically normalized fields $\hat L^\alpha= (T_L^{-1} \hat L_c)^\alpha$ gives rise to the full bosonic effective Lagrangian
\be
{\cal L}_{full} = \Big[ \big(\tilde g_{\alpha \bar \beta}+O(\epsilon) \big)  \partial  L^\alpha \partial \bar L^{\bar \beta}  \Big]- 
\Big[ \epsilon^2 \tilde g^{\bar \alpha\alpha} \overline F_{\bar \alpha,1} F_{\alpha,1} +O(\epsilon^3) \Big]\,,
\label{Lfull}
\ee
which precisely agrees with ${\cal L}_{sim}$, as given by eq.(\ref{Lsim}), at leading order in $\epsilon$.

It is important to remark that  eqs.(\ref{Fiup}) and (\ref{FiFarel}) hold also when $\hat L_c^\alpha\neq 0$, because at $O(\epsilon)$ $L^\alpha$ is undetermined. This implies that light field fluctuations arise
also from the expansion of $H^i$ at $O(\epsilon)$ from the field displacements $H_1^i$ 
in eq.(\ref{PhiH1x}).
Alternatively, we could have proceeded by just using eq.(\ref{PhiH1}), valid at the VEV level only,
and then expand in a more conventional way $H^i = \langle H^i \rangle + \hat H^i$. The terms linear in $\hat H^i$ at $O(\epsilon)$ do not vanish any more and the integration over the heavy fields has to be performed. Up to $O(\epsilon^2)$ in the potential, such integration does not present
any problem and can straightforwardly be performed, leading to eq.(\ref{Lfull}), in complete agreement with our alternative approach.


\section{Supergravity}

The analysis of a general SUGRA model 
closely follows along the lines 
of the flat SUSY case considered in the previous section, with one extra important requirement,
which is a consequence of the universal nature of the gravitational interactions.

The superpotential is taken as in eq.(\ref{WSUSY}) and all the assumptions of the flat case
continue to hold here. The SUGRA scalar potential is a generalization of eq.(\ref{VSUSY}), which reads
\be
V = e^{K/M_p^2} \bigg( g^{\bar M N} \overline F_{\bar M} F_N - 3 \frac{|W|^2}{M_p^2} \bigg) =M_p^2  e^{G/M_p^2} 
\bigg(g^{\bar M N} \overline G_{\bar M} G_N - 3 M_p^2 \bigg)\,,
\label{Vsugra}
\ee
where, as usual, $F_M=D_M W = \partial_M W +(\partial_M K) W/M_p^2$ is the K\"ahler covariant derivative
and we have also introduced the K\"ahler invariant function $G = K + M_p^2 \log( |W|^2/M_p^6)$ and its derivatives
$G_M = \partial_M G =M_p^2  F_M/W$.
$M_p$ is the reduced Planck mass that for simplicity will be set to one in what follows.
As remarked above, gravity makes the decoupling of the heavy fields from the light ones harder.
This is best seen if we expand the scalar potential in $\epsilon$ and analyze the leading term $V_0$ which reads
\be
V_0 =  e^{K} \Big( g^{\bar M N} \overline F_{0,\bar M} F_{0,N} - 3 |W_0|^2 \Big) \,,
\label{V0sugra}
\ee
where 
\be
F_{0,i} = \partial_i W_0 + (\partial_i K) W_0 \,, \ \ \ \ \ \ 
F_{0,\alpha} = (\partial_\alpha K) W_0 \,.
\label{Fsugra}
\ee
It is immediately clear from eqs.(\ref{V0sugra}) and (\ref{Fsugra})  that even at $O(\epsilon^0)$ there is in general no decoupling between the fields $H^i$ and $L^\alpha$, as was the case for flat space.  As we mentioned in the introduction and can be seen from eqs.(\ref{Fsugra}), $F_{0,i}=0$ is independent of $L^\alpha$  if either $\langle W_0\rangle \sim O(\epsilon)$ or $K$ is factorizable at leading order. We now separately discuss the two situations.

\subsection{Small $\langle W_0\rangle $}

Aside from decoupling, the most important reason to have a small  $\langle W_0\rangle $ is
the requirement of a sufficiently light gravitino mass and to ensure a mass hierarchy between the heavy and light fields. 
When $\langle W_0\rangle\sim O(\epsilon)$, the expansion of the $G_M$'s is taken as follows: 
\be
G_i  =  G_{i,_-1} + G_{i,0} \,,  \ \ \ \ G_\alpha  = G_{\alpha,0}\,,
\ee 
where we count the powers of $\epsilon$ by taking into account the presence of $W$ in the denominator of the $G$ factors.  In this way terms of $O(\epsilon^{-1})$ appear, but no poles in $\epsilon$ arise, being $e^G$ of $O(\epsilon^2)$.  
In the particular ``K\"ahler gauge" in which we defined the theory, we have
 \bea
 G_{i,-1} & = &  \frac{\partial_i W_0}{W} \,, \ \ \ G_{i,0} = \partial_i K +  \frac{\partial_i W_1}{W}\,, \nn \\
 G_{\alpha,0} & = &  \partial_\alpha K +  \frac{\partial_\alpha W_1}{W}\,.
 \eea
As in the previous flat space analysis, let us first show that the location of the vacuum is reliably computed in the simple theory,
namely that $\partial_\alpha V_{sim} = \partial_\alpha V+O(\epsilon^3)$, where $V_{sim}$ is the simple effective potential obtained
by setting $H^i=H_0^i$ in $K$ and $W$.  The full e.o.m. for the fields can be written as 
\be
\partial_M V = G_M V + e^G (G^{P}\nabla_M G_P + G_M)  = 0\,.
\label{eomGsugra}
\ee
In eq.(\ref{eomGsugra}), $G^P = \overline G_{\bar P} \, g^{\bar P P }$, $\nabla_M G_P = \partial_M G_P - \Gamma_{MP}^Q G_Q$ is the K\"ahler covariant derivative and
$\Gamma_{MP}^Q= (\partial_M g_{P\bar Q})g^{\bar Q Q}$ are the holomorphic components of the affine connection.
The expansion of  $\partial_M G_P$ gives
\bea
\partial_j G_i & = &   (\partial_j G_i)_{-2} + (\partial_j G_i)_{-1} + (\partial_j G_i)_0\,,   \  \  \ \partial_\beta G_i =(\partial_\beta G_i)_{-1} +(\partial_\beta G_i)_0\,, \nn \\
\partial_j G_\alpha&  = &   (\partial_j G_\alpha)_{-1} + (\partial_j G_\alpha)_0\,, \hspace{2.1cm}
 \partial_\beta G_\alpha =(\partial_\beta G_\alpha)_{0} \,.
\eea
For simplicity, we do not write the explicit forms of the derivatives of the $G$'s in terms of $K$ and $W$, 
being straightforward to derive these expressions.  At $O(\epsilon^0)$, $(\partial_i V)_0 = 0$ is satisfied by $(\partial_j G_i)_{-2} = - (G_i)_{-1} (G_j)_{-1} = 0$, i.e. $\partial_i W_0 = 0$, which fix $H_0^i$. The equations $(\partial_\beta V)_0 = (\partial_\beta V)_1=0$ automatically
vanish for $H^i=H_0^i$. The equations $(\partial_j V)_1=0$ give the linear order displacement of the heavy fields
(the analogue of the flat space formula (\ref{PhiH1})):
\be
H_1^i = - (\hat K^{-1})^i_{\bar j} (\overline G^{\bar j})_0\,,
\label{PhiH1sugra}
\ee
with $\hat K^{\bar i}_j  = g^{\bar i k}(\partial_k  G_j)_{-1}$ and $(\overline G^{\bar j})_0 = (\overline G_{\bar M})_0 g^{\bar M i}$, evaluated at $H^i=H_0^i$.

At the shifted vacuum $H_0^i+\epsilon H_1^i$ we have
\bea
G_i & =  &  O(1), \ \hspace{0.8cm}  G_\alpha = O(1), \hspace{0.6cm} G^i = O(\epsilon), \ \ \  G^\alpha = O(1)\,, \nn \\
\nabla_\beta G_i & = &   O(1), \ \ \ \nabla_\beta G_\alpha = O(1),\ \ \ \ \ \ V= O(\epsilon^2) \,. \label{GdG} 
\eea
Using eq.(\ref{MatrixId}), we also have
\bea
G^\alpha & = &  \overline G_{\bar \alpha}\tilde  g^{\bar \alpha \alpha} + O(\epsilon) \,, \nn \\
\nabla_\beta G_\alpha & = &  \partial_\beta G_\alpha - \tilde g^{\bar \gamma \gamma} \partial_\beta g_{\alpha\bar \gamma} G_\gamma = (\tilde \nabla_\beta G_\alpha) + O(\epsilon)\,,
\label{GnablaG}
\eea
where $\tilde \nabla$ is the covariant derivative constructed in the subspace parametrized by the scalar fields $L^\alpha$ only, namely the one entering in $V_{sim}$. Finally, using eqs.(\ref{GdG}) and (\ref{GnablaG}), it follows that $\partial_\beta V = \partial_\beta V_{sim} + O(\epsilon^3)$, where
\be
V_{sim} =  e^{G}  \bigg[\tilde g^{\bar \alpha \alpha}\overline G_{\bar\alpha } G_\alpha - 3\bigg]\,,
\ee
with heavy moduli frozen at $H_0^i$.

The equivalence of the simple low-energy effective theory with the full one proceeds along the same lines
of the flat space SUSY case discussed before.  
The expansion of the scalar potential (\ref{Vsugra})  is as follows: $V=V_0+\epsilon V_1+\epsilon^2 V_2$, with
\bea
V_0 & = & e^G g^{\bar i j}\overline G_{\bar i,-1} G_{j,-1}\,, \nn \\
V_1 & = & e^G \Big[g^{\bar i j}\overline G_{\bar j,-1} G_{j,0}+g^{\bar j \alpha}\overline G_{\bar j,-1} G_{\alpha,0}+c.c \Big] \,,
\label{V012sugra} \\
V_2 & = & e^G \bigg[g^{\bar i j}\overline G_{\bar j,0} G_{j,0}+ g^{\bar \alpha \alpha}\overline G_{\bar \alpha,0} G_{\alpha,0}+ \Big(g^{\bar i \alpha}\overline G_{\bar i,0} G_{\alpha,0}+ c.c\Big)-3\bigg]\,. \nn
\eea
The canonically normalized field fluctuations are $\hat \phi_c^M$, with $\hat\phi = T \hat \phi_c$,  $\hat{\bar\phi} = T^* \hat{\bar  \phi}_c$ and $g_0 =  (T^{-1})^\dagger T^{-1} $. The  $\hat H_c^i$ are still the proper heavy field fluctuations, as is found by expanding the leading potential term $V_0$ at quadratic order. The leading mass term, in the parametrization (\ref{WSUSY}), is given by $\exp(K)  T_H^\dagger M^2 T_H$, with $M^2$ as in eq.(\ref{Mass0}). In total analogy with the flat case, the integration of  $\hat H_c^i$ is trivial, just fixing
$\hat H_c^i=0$.  One has
$G_{i,-1}=0$ and again $G^i_0=0$ so that 
\be
V  =e^G   \Big[ \tilde g^{\bar \alpha \alpha}  \overline G_{0,\bar \alpha} G_{0,\alpha} - 3\Big]\,.
\ee
Going back to non-canonically normalized fields $\hat L^\alpha= (T_L^{-1} \hat L_c)^\alpha$ gives rise to the full bosonic effective Lagrangian
\be
{\cal L}_{full} = \Big[ \big(\tilde g_{\alpha \bar \beta}+O(\epsilon) \big)  \partial L^\alpha \partial \bar L^{\bar \beta}  \Big]- 
 \bigg[ e^G\Big( \tilde g^{\bar \alpha\alpha} \overline G_{0,\bar \alpha} G_{0,\alpha}-3\Big) +O(\epsilon^3) \bigg]\,,
\ee
which agrees with  ${\cal L}_{sim}=\tilde  g_{\alpha \bar \beta}  \partial L^\alpha \partial \bar L^{\bar \beta}-V_{sim}$, at leading order in $\epsilon$.

\subsection{Almost Factorizable K\"ahler Potential}

A decoupling between the heavy and light fields is possible also for generic $\langle W_0 \rangle$, provided that
$K$ is almost factorizable, namely
\be
K(\phi,\bar \phi) = K_H(H^i,\bar H^{\bar i}) +  K_L(L^\alpha,\bar L^{\bar \alpha}) + \epsilon  K_{mix}(\phi^M,\bar \phi^{\bar M})\,.
\label{FactorK}
\ee 
With $K$ as in eq.(\ref{FactorK}),  the SUSY equations $F_{0,i}=0$  do not depend on $L^\alpha$ and lead to solutions of the e.o.m.  $(\partial_i V)_0=0$ on a generic non-SUSY vacuum with  $F_{0,\alpha}\neq 0$. This is expected, since a superpotential of the form (\ref{WSUSY}) at $O(\epsilon^0)$ does not depend on $L$ and hence the sufficient conditions for  decoupling discussed in \cite{Binetruy:2004hh}, namely $K = K_H(H^i,\bar H^i) + 
K_L(L^\alpha,\bar L^\alpha)$ and $W = W_H(H^i) W_L(L^\alpha)$ are fulfilled at $O(\epsilon^0)$, with $W_H(H)=W_0(H)$
and $W_L(L)=1$. As we will see, showing how the decoupling works for a factorizable $K$ is much simpler than the
small $W_0$ case, since it is enough to work out the scalar potential at $O(\epsilon^0)$.  The latter reads
\be
V_0 = e^{G_0} (G^i_0 G_{i ,0}+ G^\alpha_0 G_{\alpha,0} - 3)\,,
\label{V0Factor}
\ee
with
\be
G_0 = K_H +K_L + \log|W_0|^2
\ee
and $G_{M,0} = \partial_M G_0$. The SUSY conditions $G_{i,0}=0$ do not depend on $L^\alpha$ and 
since $\partial_{\alpha} G_{i,0} = \partial_i G_{\alpha,0} = 0$, they automatically are solutions of the e.o.m.
$\partial_i V_0=0$, independently of $L^\alpha$ and of $G_{\alpha,0}$.
The $O(\epsilon^0)$ e.o.m. for $L^\alpha$ are not necessarily trivial as in the small $W_0$ situation, since 
$\partial_\alpha V_0=0$ do not automatically vanish. In this case, the vacuum is entirely determined by $V_0$, which 
effectively encodes the leading coupling terms between $H^i$ and $L^\alpha$.
The expansion of $V_0$ in heavy field fluctuations is of the schematic form
(\ref{SchematicExp}). It is straightforward  to see from eq.(\ref{V0Factor})  that all the mixed derivatives
$\partial_i \partial_{\alpha_1}\ldots \partial_{\alpha_n} V_0$ vanish \cite{Achucarro:2007qa}. As a result,  
$\partial_H V$ is at most of  $O(\epsilon)$, $\partial_H^2 V$ is of $O(1)$, implying that $\hat H_c$ is  $O(\epsilon)$.
The leading bosonic effective Lagrangian at $O(\epsilon^0)$ is hence reliably
determined by just freezing the heavy fields to their VEV's $H_0^i$. 

It can happen that $\partial_\alpha V_0$ is trivial, if the SUGRA model is of the no-scale type, with
\be
G^{\alpha}_0 G_{\alpha,0} = g_L^{\bar \alpha \alpha} \partial_{\bar \alpha} K_L \partial_\alpha K_L = 3\,,
\label{noscale}
\ee
with $g^L$ the inverse of the K\"ahler metric $\partial_{\alpha}\partial_{\bar\alpha} K_L$. When eq.(\ref{noscale}) is valid,
eq.(\ref{V0Factor}) simplifies to
\be
V_0 = e^{G_0} G^i_0 G_{i,0}\,,
\ee
and $\partial_\alpha V_0$ automatically vanish for $G_{i,0}=0$. The situation is now very similar to the non-SUSY
case discussed in section 2. The fields $L^\alpha$ are generally fixed by the $O(\epsilon)$ e.o.m. $\partial_\alpha V_1=0$,
where $V_1$ is the next scalar potential term, whose explicit form will not be needed. The expansion in fluctuations
of $V_0$ is as before, with $\hat H_c \sim O(\epsilon)$.  The leading bosonic effective Lagrangian arises at $O(\epsilon)$ 
and is determined by just freezing the heavy fields to their VEV's $H_0^i$. As we will see in Appendix C,
the large volume models of \cite{Balasubramanian:2005zx,Conlon:2005ki} belong to this class of models.

\section{Conclusions}

Our results show once again the power of supersymmetry combined with an effective field theory approach,
allowing an enormous simplification in the description of complicated field theories with many fields, as the ones
appearing in the string landscape. Probably the most important message of our paper is that the often assumed restriction of factorisable K\"ahler potentials, required to disentangle heavy and light degrees of freedom for superpotentials of the form (\ref{Wint}), can be relaxed and replaced by the single condition of a small $W$ at the vacuum (when gravity is included). 
Unless the K\"ahler potential is abnormally large, as in the large volume models of \cite{Balasubramanian:2005zx,Conlon:2005ki}, the smallness of $W$ is phenomenologically required to have a light gravitino, i.e. low-energy SUSY breaking. 
Our main working assumption is eq.(\ref{Wint}),
which essentially imply that the light fields are stabilized by a dynamics parametrically
suppressed compared to the dynamics responsible for the heavy fields stabilization. 
In the context of effective SUGRA theories with no gauge fields, this assumption is quite
natural, ensuring a mass gap between the spectrum of heavy and light fields. This is in fact
a common setup in string compactifications with fluxes,  where some moduli are stabilized by
flux generated superpotentials with relatively strong couplings, the remaining moduli feeling only
much weaker non-perturbative generated couplings. We have mentioned in the introduction how our results
provide a solid framework for the way complex structure and dilaton fields are treated in KKLT--like models
and how the fine-tuning of $W_0$ is actually more important than previously thought.\footnote{The smallness of $W_0$ is not necessarily obtained by fine-tuning in theories with an approximate R-symmetry or R-parity, where  
$W_0$ can naturally be small. It would be very interesting to find string models with this property.}
It is obvious that all these reasonings apply to any other string (or non-string) derived SUGRA model with
a superpotential of the form (\ref{Wint}). 
For simplicity, we have considered in this paper field theories
with a single hierarchy in the field mass spectrum, but generalizations to multiple hierarchies should be straightforward.

The most compelling generalization of our results is the introduction of gauge fields and charged matter
and see under what conditions one can still reliably freeze moduli in this context.
The issue is particularly important in so called string local models,  where all moduli are assumed to be frozen
and gravity decoupled, and one constructs (semi)-realistic string models using essentially
the much simpler and general model building techniques valid for non-compact spaces (see e.g.\cite{Beasley:2008dc} for an interesting class of F-theory models of this kind). We hope to come back to this important point in a forthcoming paper \cite{EffD}.

\section*{Acknowledgments}

We would like to thank S. Cecotti and S. Kachru for useful comments on the manuscript, S.P. de Alwis
for helpful clarifications on his works and especially A. Romanino for an enlightening discussion.
This work has been partially supported by the European Community's Human
Potential Programme under contracts MRTN-CT-2004-005104,
by the Italian MIUR under contract PRIN-2005023102 and by INFN.

\appendix

\section{A Flat Space Model}
 
We show here how the general analysis performed in section 3 works in 
a concrete example with global ${\cal N}=1$ SUSY. For simplicity, we consider just two fields.
The K\"ahler and superpotential terms are 
\bea
K  & = & \bar H H +\bar L L -\frac{(\bar L L )^2}{4\Lambda^2} + \eta (\bar H L + \bar L H) \,,\nn \\
W & = &  -\frac M 2 H^2 + \frac{H^3}{3} + \epsilon \Big( \mu^2 L + \frac Y2 H L^2\Big)\,,
\eea 
with all parameters taken to be real. Positive definiteness of the metric requires $\eta<1$.
When $H$ is frozen, the model is a simple deformation of the Polonyi model which admits non-SUSY vacua.
Due to the $R$--symmetry breaking $HL^2$ operator, the non-SUSY vacuum is displaced from the origin in $L$. The heavy field $H$ enters in $W_1$ and has a mixing term of $O(1)$
with $L$ due to the last term in $K$. We take $H_0=\bar H_0 = M$ as solution to the SUSY equation $F_{H,0}=0$. Neglecting irrelevant terms, the simple effective K\"ahler and superpotential terms for $L$ are given by 
\be
K_{sim} = \bar L L -\frac{(\bar L L )^2}{4\Lambda^2} \,, \ \ \ \ \ W_{sim} = \epsilon \Big( \mu^2 L + \frac Y2 M L^2\Big)\,,
\ee
resulting in the simple bosonic low-energy Lagrangian
\be
{\cal L}_{sim} = \Big(1- \frac{|L|^2}{\Lambda^2}\Big) |\partial L|^2 - \epsilon^2 \frac{1}{1- |L|^2/\Lambda^2} |\mu^2 + Y M L|^2\,.
\label{LeffExp}
\ee
The non-SUSY vacuum of $V_{sim}$ is easily found to be 
\be
L_0 = -\frac{Y  M \Lambda^2}{\mu^2}\,.
\label{L0vevexp}
\ee
It is a simple exercise to show that the vacuum $(H_0,L_0)$ is a (meta-stable) minimum with
\bea
m^2_L & = & m^2_{\bar L} =  \epsilon^2 \frac{\mu^{12}}{\Lambda^2 c^2}\,, \nn \\
F_L&  = &  \epsilon \frac{c}{\mu^2} \,, \ \ \ \ \ \ F^L = \epsilon \mu^2\,,
\label{mFL}
\eea
where
\be
c\equiv  \mu^4 - Y^2 M^2 \Lambda^2\,.
\ee
 
We now show that the same effective Lagrangian (\ref{LeffExp}) is obtained by properly integrating out the canonical
heavy degrees of freedom.  We start by recomputing the vacuum of the full potential in a series expansion in $\epsilon$.
At leading order $H_0 = \bar H_0 = M$. At next order, one easily finds
\be
H_1(x) = -\frac{YL(x)^2}{2M} + \frac{\eta(\mu^2 +  L(x) MY)} {M\big(1-\frac{|L|^2(x)}{\Lambda^2}\big)}\,.
\label{H1exp}
\ee
We have written the space-time dependence of the fields in eq.(\ref{H1exp}) to emphasize that
this formula is not only valid at the (constant) vacuum but for any light field fluctuation, being $L$ undetermined at this order. 
At $O(\epsilon^2)$ one gets the same equations of motion for $L$ found with $V_{sim}$, 
as well as the shift $H_2$, the explicit form of which is not necessary. The leading order vacuum $(H_0,L_0)$ coincides then with that obtained before. We can expand the full potential $V$ in small fluctuations $\hat \phi=(\hat H,\hat L)^t$ by writing $H(x) = H_0+\epsilon H_1(x) + \hat H(x)$, $L(x) = L_0 + \hat L(x)$. The $2\times 2$ metric $g_0=g(H_0,L_0)$ can be written as $g_0 = (T^{-1})^\dagger T^{-1}$, with $T$ the lower triangular matrix
\be
T = \left(
\begin{matrix}
\sqrt{\frac{c}{c-\eta^2 \mu^4}} &  0  \cr
\frac{\eta \mu^4}{\sqrt{c(c-\eta^2\mu^4)}} & \frac{\mu^2}{\sqrt{c}}  \cr
\end{matrix}\;
\right)\;,\label{Utr}
\ee
valid for $\mu^2> Y M \Lambda$. The canonical Lagrangian is obtained by writing $\hat \phi = T\hat \phi_c$.
After some algebra, it is not difficult to see that the canonically normalized fields $\hat H_c$ are also mass eigenvectors
with a mass given, at leading order, by
\be
m^2_H = m^2_{\bar H} = M^2 \frac{c^2}{(c-\eta^2\mu^4)^2} + O(\epsilon)\,.
\ee
The $F$-terms for $L$ are as in eq.(\ref{mFL}) whereas for the heavy field we have
\be
F_H = \epsilon \eta \mu^2\,, \ \ \ \ \ \ F^H = O(\epsilon^2)\,,
\ee
as expected from our general analysis. Setting $\hat H_c = \bar{\hat H}_c = 0$ gives the full bosonic effective Lagrangian ${\cal L}_{full}$ for the light fluctuations. It is now a simple, yet algebraically lengthy, exercise to show that at leading order  ${\cal L}_{full}$ coincides with ${\cal L}_{sim}$.

 \section{Fine Tuned $W$: a KKLT--like Model with Broken SUSY}

In this appendix we apply the results of the paper to a KKLT--like SUGRA toy model in a non-SUSY vacuum with $n_H=n_L=2$. Although the model is pretty simple and contains just four complex fields, instead of hundreds as in realistic string models,
it is already sufficiently complicated to make an analytical study a formidable task. For this reason we opt here
for a numerical analysis. For obvious reasons, we will not show the equivalence of the entire effective potentials, but
we will restrict our attention on the comparison of the vacuum and the scalar mass spectrum. The K\"ahler and superpotential terms are taken as follows:
\bea
\hspace{-0.9cm} K & = & -  \log\Big[ (T+\bar T)^{3/2} + \xi(S+\bar S)^{3/2}\Big]^2(Z+\bar Z)(S+\bar S) + \frac{\bar U U}{(Z+\bar Z)^2} -  \eta \frac{(\bar U U)^2}{(Z+\bar Z)^4} \,, \label{K4sugra} \\
\hspace{-0.9cm} W & = & a Z^2 + b Z + S (c Z^2 + d Z + e) + Y U Z^2 + \beta Z^2 e^{-\alpha T} \,. \label{W4sugra}
\eea
The heavy fields are $S$ and $Z$, the light ones $T$ and $U$. The fields $S$ and $Z$ mimic respectively the dilaton and a complex structure modulus of some IIB Calabi-Yau compactification, $T$ represents the overall universal K\"ahler modulus and $U$ a generic Polonyi field responsible for SUSY breaking. The kinetic mixing between heavy and light fields
is provided by the universal $\alpha^\prime$ correction to the volume \cite{Becker:2002nn}, parametrized by $\xi$ in eq.(\ref{K4sugra}) and by the complex structure dependence of the kinetic term for $U$.
The first five terms in the superpotential mimics superpotential terms arising from fluxes and should be 
identified with $W_0$ in our general analysis (see e.g. eq.(3.53) of \cite{Lust:2005dy}).
The term $Y U Z^2$ is a cubic coupling responsible for the main SUSY breaking source in the $U$ direction
and $Z^2 \exp(-\alpha T)$ is a non-perturbatively generated coupling, responsible for the stabilization of $T$, which 
might arise from euclidean $D_3$--instantons or gaugino condensation from $D_7$--branes. These are the small
superpotential terms defining $W_1$. The prefactor of the exponential is a function of the complex structure moduli, which for simplicity we have taken to be just quadratic. It should be stressed that the detailed structure of $K$ and $W$ in eqs.(\ref{K4sugra}) and (\ref{W4sugra}) is pretty arbitrary and not well motivated; it should just be seen as a simple, yet not trivial, string--inspired example where to concretely apply our results. 
The parameters entering in eqs.(\ref{K4sugra}) and (\ref{W4sugra}) have been partially fixed by requiring to have a small positive cosmological constant, $O(1)$ TeV gravitino mass (which fixes the size of $W_0$ to be $10^{-11}$), $S_0\simeq Z_0 \simeq 10$, small $U_0$ and $T_0\sim O(10)$.  None of these requirements clearly affect the results of our analysis, but they make our example more ``realistic''.
We have taken\footnote{In a realistic string setting, the fluxes are quantized and makes the appearance of terms like $a$ and $b$ in eq.(\ref{explPARA}) rather difficult, if not impossible. 
We do not care about this possible issue. As explained above, we do not pretend our model to mimic a realistic string set-up
in any single detail.}   
\bea
a &  = &  -2.55+10^{-13}\,, \ \ \ b = 25.5+10^{-12}\,, \ \ \  c = 0.25\,, \ \ \ d = -2.45\,, \ \ \ e=-0.5 \nn \\
\xi & = & 1\,, \ \ \ \eta = 100\,, \ \ \ \alpha = 1\,, \ \ \ \beta = - 10^{-4}\,, \ \ \ Y = 0.83\cdot 10^{-14}\,.
\label{explPARA}
\eea
The smallness of $Y$, required to get a dS vacuum with a sufficiently small cosmological constant, justifies
the location of the $U Z^2$ coupling in $W_1$. The SUSY VEV's of $S$ and $Z$, as given by $\partial_S W_0 = \partial_Z W_0$ are precisely $S_0 = Z_0 = 10$,
with $W_0(S_0,Z_0) = 10^{-11}$. In table 1 we report the exact VEV's and $F$--terms of the fields as numerically computed in the full model and their relative shifts compared to those computed, again numerically, in the simple effective model, with $S$ and $Z$ frozen at their values $S_0$ and $Z_0$.  The model above belongs to the general class of models studied in the main text with an  $\epsilon$ roughly  $O(10^{-12})$. Keeping two significant digits, the physical masses are
\bea
m_{H1}^2 & = &  2.9\cdot 10^{-2},  \ \ m_{H2}^2 = 2.7\cdot 10^{-2}, \\
m_{T_i}^2 & = & 6.6\cdot 10^{-27}, \ \  
m_{T_r}^2 = 6.2\cdot 10^{-27}, \ \ 
m_{U_r}^2 = 1.6\cdot 10^{-27}, \ \  
m_{U_i}^2 = 1.6\cdot 10^{-27}, \nn
\eea
where $H_1\simeq Z-S$, $H_2\simeq Z+S$ and the subscript $r$ and $i$ denote real and imaginary field components, respectively. The masses $m_{H1}^2$ and $m_{H2}^2$ refer to both components of the complex scalar fields, being SUSY breaking effects negligible. The mass shifts are
\be
\frac{\Delta m_{T_i}^2}{m_{T_i}^2} = 3 \cdot 10^{-14}\,, \ \ 
\frac{\Delta m_{T_r}^2}{m_{T_r}^2} = 2 \cdot 10^{-14}\,, \ \ 
\frac{\Delta m_{U_i}^2}{m_{U_i}^2} = 9 \cdot 10^{-14}\,, \ \ 
\frac{\Delta m_{U_r}^2}{m_{U_r}^2} =  10^{-13}\,.
\ee
Finally, we also report the gravitino mass, the cosmological constant and their relative shifts:
\bea
m_{3/2}^2 &  = &  1.3\cdot 10^{-30} \,, \ \ \  \frac{\Delta m_{3/2}^2}{m_{3/2}^2} =-4 \cdot 10^{-14}\,, \nn \\
V_0 & = &  2.7 \cdot 10^{-32}\,, \ \ \ \frac{\Delta V_0}{V_0} = 2\cdot 10^{-11}\,.
\eea
Notice that being the cosmological constant fine-tuned to be ``small", namely of order  $10^{-2} m_{3/2}^2$,
its relative shift is larger. The latter is inversely proportional to the smallness of $V_0$.

The values of the $F$--terms  agree pretty well with the scalings in $\epsilon$ as expected from our general analysis.
The relative shifts are typically smaller than $\epsilon$, due to the fact that the K\"ahler mixing between the $H$ and $L$ fields
coming from  (\ref{K4sugra}) in the above vacuum are relatively small.  This example shows the excellent agreement between the full and the simple theory.  

\TABLE[t]{
\begin{tabular}{|c|c|c|c|c|c|c| }
\hline&
$\langle X \rangle $ & $\Delta \langle X \rangle/\langle X \rangle$ &$F^X$ &$\Delta F^X/F^X$ \\ \hline
$S$ & $10+2\cdot 10^{-13}$ & $2\cdot 10^{-14}$ &  $8\cdot 10^{-24}$ & --  \\
$Z$ & $10+3\cdot 10^{-13}$ &   $3\cdot 10^{-14}$ & $-3\cdot 10^{-23}$ & -- \\
$T$ & $23.8$ &   $4\cdot 10^{-15}$  & $-3\cdot 10^{-11}$ &  $8\cdot 10^{-14}$  \\
$U$ &  $0.05$ &  $-3\cdot 10^{-13}$  & $3\cdot 10^{-10}$ &  $ 10^{-13}$ \\
 \hline
\end{tabular}
\caption{VEV's and $F$--terms for the fields and their relative shifts, as derived by a numerical analysis. Here and in the main text $\Delta X /X \equiv (X_{full}-X_{sim})/X_{full}$. All quantities are in reduced Planck units.}
}
 
\section{Factorizable K\"ahler Potential: a Large Volume Model}

Type IIB SUGRA compactifications on CY admit  vacua where the internal volume is very large,
resulting in so called large volume models \cite{Balasubramanian:2005zx,Conlon:2005ki}. 
A decoupling occurs in these models, since they approximately satisfy the
conditions for decoupling given in \cite{Binetruy:2004hh}.\footnote{It has been already noted in Ref.~\cite{Achucarro:2008fk} that large volume models  satisfy an approximate condition of decoupling, but this observation has not been fully exploited there.} Unfortunately, we did not find an easy way to place such models in a general setting, so that 
we will focus on an explicit known example (``swiss-cheese'' compactification).

Consider a type IIB CY orientifold compactification on $\textbf{CP}^4_{[1,1,1,6,9 ]}$. 
This manifold has $h^{1,1} =2$ and $h^{2,1}=272$ K\"ahler and complex structure moduli, respectively. 
In order to be able to treat this system, we keep only one complex structure modulus, denoted by $Z$ in the following.
As will be clear, our conclusions do not really depend on such drastic simplification. 
In the usual adiabatic approximation of neglecting flux effects, but keeping the universal $\alpha^\prime$ correction 
\cite{Becker:2002nn}, the K\"ahler potential reads 
\be
K=-2 \log\bigg[ {\rm Vol}(T,t)+\xi(S+\bar S)^{3/2}\bigg]-\log (S+\bar S) -\log(Z+\bar Z)\,,
\label{IIBkaehler}
\ee
where ${\rm Vol}$ is the CY volume, which depends on the two K\"ahler moduli $T$ and 
$t$ \cite{Denef:2004dm}:
\be
{\rm Vol}(T,t)= \frac{1}{9\sqrt{2}}\left(T_r^{3/2}-t_r^{3/2}\right)\,.
\label{KVol}
\ee
In eq.(\ref{KVol}) we generally denoted by $X_r = (X+\bar X)/2$ the real part of a complex field $X$.
The small expansion parameter in this class of models is $\epsilon \equiv 1/{\rm Vol}\sim T_r^{-3/2}$, with $T_r\gg 1$ and $t_r\gtrsim 1$. 
The superpotential is the sum of a flux superpotential $W_0=W_0(S,Z)$ and a non--perturbative term $W_1=W_1(S,Z,t)$.\footnote{Since $T$ is very large, possible non--perturbative terms of the form $\exp(- aT)$ are totally negligible.}
 The stabilization to large volume requires $W_1 \sim \epsilon$ and hence the superpotential has the form of eq.(\ref{WSUSY}):
\be
W = W_0(S,Z) + \epsilon W_1(S,Z,t)\,,
\ee
where $\epsilon$ is introduced in $K$ by redefining $T\rightarrow \epsilon^{-2/3} T$.  
Eq. (\ref{IIBkaehler}) factorizes at $O(1)$, namely
\be
K = K_H(S,Z)+K_L(T,t) + \epsilon K_{mix}(S,T,t)\,,
\ee
with $K_H = - \log(4S_r Z_r)$ and $K_L= -2\log\,{\rm Vol}$.
Contrary to the KKLT--like models, no fine-tuning on $W_0$ is required for decoupling.
It is straightforward to expand $G_M$ in powers of $\epsilon$. We get
\bea
G_S & = & G_{S,0} + O(\epsilon)\,, \hspace{1.7cm}
G_Z =  G_{Z,0} +  O(\epsilon)\,, \nn \\
G_T & = & \epsilon^{2/3}  G_{T,2/3} + O(\epsilon^{5/3})\,, \ \ \
G_t  = \epsilon \,G_{t,1}  + O(\epsilon^2)\,,\nn
\eea
with
\bea
G_{S,0} &  = &  -\frac{1}{2S_r} +\frac{\partial_S W_0}{W_0}\,, \ \ \
G_{Z,0}   =  -\frac{1}{2Z_r} + \frac{\partial_Z W_0}{W_0}\,, \nn \\
G_{T,2/3} &  = &   -\frac{3}{2T_r}\,, \hspace{1.8cm}
G_{t,1}  =    \frac{3\sqrt{t_r}}{2T_r^{3/2}}+\frac{\partial_t W_1}{W_0}\,. \nn 
\eea
The expansion of the scalar potential $V$ is as follows: $V= \epsilon^2 V_{2}+\epsilon^3 V_{3} + O(\epsilon^4)$, where
\bea
V_2  & = &    e^G_2\Big[g^{\bar Z Z}_0 \overline G_{\bar Z,0} G_{Z,0} + g^{\bar S S}_0 \overline G_{\bar S,0} G_{S,0}  \Big] \,,  \\
e^G & = &  \epsilon^2 e^G_2 + O(\epsilon^3), \ \ \ e^G_2 =\frac{81|W_0|^2}{2 S_r Z_r T_r^3}\,.
\eea 
It is important to notice that, aside from the overall $T$--dependence appearing in $e^G_2$, $V_2$ depends on $S$ and $Z$ only. 
A possible non-trivial K\"ahler moduli dependence of $O(\epsilon^2)$ in $V_2$ exactly cancels the -3 term in $V_2$, due to the approximate no-scale structure of the K\"ahler potential, for which \cite{Covi:2008ea}
\be
\sum_{i,j=T,t}g^{\bar j i} \partial_i K \partial_{\bar j} K = 3+O(\xi \epsilon)\,.
\ee
An effective decoupling between $S$, $Z$ and the K\"ahler moduli appears at this order, so that
the leading e.o.m. for $S$ and $Z$ admit the SUSY solutions 
\be
G_{S,0}(S_0,Z_0) = G_{Z,0}(S_0,Z_0) = 0\,.
\label{GSZ0}
\ee
Around the solutions (\ref{GSZ0}), all the terms linear in the heavy field fluctuations $\hat H=\hat S,\hat T$ of the form $c_n \hat H \hat L^n$, where $\hat L=\hat T,\hat t$, $n\geq 1$, vanish. 
These terms can only arise from $V_3$, so that schematically we have $O(\epsilon^2) \hat H^2 + O(\epsilon^3) c_n \hat H \hat L^n$, implying that $\hat H \sim O(\epsilon)$. Hence integrating out $\hat H$ can only result in effective couplings
of $O(\epsilon^4)$ or higher and hence we can effectively fix $\hat H=0$. 
Rather than proceeding as in the main text, by canonically normalizing the fields and so on, 
in this particular model it turns out to be much simpler to directly compute the full leading effective potential 
$V_3$ for $T$ and $t$. We find 
\be
V_3 = \frac{27\Big[81 \xi_0 |w_0|^2+4 \sqrt{2t_r}  T_r^{3} |\partial_t w_1|^2 - 3 
t_r T_r^{3/2} (\bar w_0 \partial_t w_1 + c.c.)\Big]}{T_r^{9/2} Z_{r,0} S_{r,0}} \,,
\ee
where $\xi_0 \equiv \xi S_{r,0}^{3/2}$, $w_0= W_0(S_0, Z_0)$ and $w_1(t) = W_1(S_0, Z_0,t)$.
It is easy to check that $V_3$  precisely coincides with the simple scalar potential constructed from $W_{sim} = W(S_0,Z_0)$ and  $K_{sim} = K(S_0,Z_0)$, where the dilaton and the complex structure modulus are frozen at their SUSY values $S_0$, $Z_0$ \cite{Balasubramanian:2005zx}. 
The scaling in $\epsilon$ of the mass spectrum is easily computed. 
The complex structure modulus, the dilaton and the gravitino mass arise from $V_2$ and hence
\be
m_{3/2}\sim m_S \sim m_Z \sim \epsilon\,.
\label{massSZ}
\ee
The K\"ahler structure moduli mass matrix arises from $V_3$. 
Given the structure of the kinetic metric, it is simple to see that
\be
m_T \sim \epsilon^{3/2}\,, \ \ \ m_t \sim \epsilon\,.
\label{massTt}
\ee
The scalings (\ref{massSZ}) and (\ref{massTt}) are in agreement with the one reported in the table 1 of
 \cite{Conlon:2005ki}. Despite the absence of a hierarchy in the mass scales of the $H$ and $L$ fields, the model is effectively decoupled and the simple naive effective theory, where one freezes $S$ and $Z$, is reliable.

\end{document}